\documentclass[10pt,conference,letterpaper]{IEEEtran}

\ifCLASSINFOpdf
\else
\fi

\usepackage{CJK}
\usepackage{amsmath}
\usepackage{indentfirst}
\usepackage{url,cite}
\usepackage{color}
\usepackage{amsfonts}
\usepackage{graphicx}
\usepackage{float}
\usepackage{booktabs}
\usepackage{multirow}
\usepackage{clrscode}
\usepackage{algorithm}
\usepackage{algorithmic}
\usepackage{comment}
\usepackage{epstopdf}
\usepackage{lscape}
\usepackage{commath}
\usepackage{diagbox}

\floatname{algorithm}{Procedure}

\usepackage[load=prefixed]{siunitx}
\newsavebox{\measurebox}

\ifCLASSOPTIONcompsoc
\usepackage[caption=false,font=normalsize,labelfont=sf,textfont=sf]{subfig}
\else
\usepackage[caption=false,font=footnotesize]{subfig}
\fi

\graphicspath{{fig/}}

\begin{document}

\title{RF-Rhythm: Secure and Usable Two-Factor RFID Authentication}

\author{
	\IEEEauthorblockN{Jiawei Li\IEEEauthorrefmark{1}, Chuyu Wang\IEEEauthorrefmark{2}, Ang Li\IEEEauthorrefmark{1}, Dianqi Han\IEEEauthorrefmark{1}, Yan Zhang\IEEEauthorrefmark{1}, Jinhang Zuo\IEEEauthorrefmark{3}, Rui Zhang\IEEEauthorrefmark{4}, Lei Xie\IEEEauthorrefmark{2}, \\ Yanchao Zhang\IEEEauthorrefmark{1}}
	\IEEEauthorblockA{\IEEEauthorrefmark{1} Arizona State University, \IEEEauthorrefmark{2} Nanjing University, \IEEEauthorrefmark{3} Carnegie Mellon University, \IEEEauthorrefmark{4} University of Delaware\\
	\{jwli, anglee, dqhan, yanzhangyz, yczhang\}@asu.edu, wangcyu217@gmail.com, jzuo@andrew.cmu.edu, ruizhang@udel.edu, \\ lxie@nju.edu.cn}
}

\maketitle

\begin{abstract}
Passive RFID technology is widely used in user authentication and access control. We propose RF-Rhythm, a secure and usable two-factor RFID authentication system with strong resilience to lost/stolen/cloned RFID cards. In RF-Rhythm, each legitimate user performs a sequence of taps on his/her RFID card according to a self-chosen secret melody. Such rhythmic taps can induce phase changes in the backscattered signals, which the RFID reader can detect to recover the user's tapping rhythm. In addition to verifying the RFID card's identification information as usual, the backend server compares the extracted tapping rhythm with what it acquires in the user enrollment phase. The user passes authentication checks if and only if both verifications succeed. We also propose a novel phase-hopping protocol in which the RFID reader emits Continuous Wave (CW) with random phases for extracting the user's secret tapping rhythm. Our protocol can prevent a capable adversary from extracting and then replaying a legitimate tapping rhythm from sniffed RFID signals. Comprehensive user experiments confirm the high security and usability of RF-Rhythm with false-positive and false-negative rates close to zero.  
\end{abstract}

\section{Introduction}
Passive (battery-less) RFID technology has been widely used in user authentication and access control. An RFID system consists of a backend server, RFID readers, and RFID cards (tags). An RFID reader sends wireless signals to interrogate a nearby RFID card, which returns its identification information by backscattering the reader's signals. The RFID reader then forwards the received information to the backend server for comparison with the stored information. If a match is found, the RFID user passes authentication and is permitted to access critical resources or enter a protected area such as a business building, parking garage, car, or even home.

Lost/stolen/cloned RFID cards pose the most critical threat to RFID authentication systems. In particular, RFID cards are often of small size and can be easily lost or stolen; they can also be cloned with many cheap existing tools. Since RFID cards are not password-protected, the adversary can use a lost/stolen/cloned RFID card to pass authentication and impersonate the legitimate user. An effective countermeasure can be two-factor authentication which requires the RFID user to present the second piece of identification information. One such solution requires the RFID user to additionally input a PIN code on a keypad \cite{RFID2FA}. It not only diminishes the convenience of contactless RFID authentication but also requires a nontrivial infrastructure update to existing RFID systems. Another plausible solution is exploring commercial mobile 2FA solutions such as Duo Mobile \cite{Duo}, which require the RFID user to manually acknowledge an authentication request on his/her enrolled smartphone. This solution needs the RFID user to own and always carry a smartphone with good network connectivity, which may not be feasible in practice. 

We propose RF-Rhythm, a secure and usable two-factor RFID authentication system with strong resilience to lost/stolen/cloned RFID cards. In RF-Rhythm, each legitimate user performs a sequence of taps on his/her RFID card according to a self-chosen secret melody. Such rhythmic taps can induce phase changes in the backscattered signals, which the RFID reader can detect to recover the user's rhythm. In addition to verifying the RFID card's identification information as usual, the backend server compares the recovered rhythm with what it acquires in the user enrollment phase. The user passes authentication only if both verifications succeed. 

The security, usability, and feasibility of RF-Rhythm lie in many aspects. First, a user can easily select a secret yet familiar song segment which is very difficult for others to guess. Second, different users may interpret the same song segment in various ways, resulting in diverse rhythmic tap patterns on the card. This means that even if the adversary knows the secret song segment, it may still have great difficulty performing the correct tapping rhythm on the RFID card. Third, RF-Rhythm is naturally resilient to traditional replay and relay attacks on RFID authentication systems. Fourth, the phase information of backscattered signals is readily available on commercial RFID readers, so RF-Rhythm only needs a minor software update to the RFID reader and backend system. Finally, RF-Rhythm applies to COTS RFID cards and does not need the user to carry any other device.

Although rhythm-based authentication has been proposed for smartphones \cite{ChenYou15} and smartwatches \cite{HutchBea18}, we are the first to explore it in RFID systems and face two unique challenges.  

\textbf{The first challenge is rhythm detection and classification, i.e., how to detect and verify the tapping rhythm from noisy RFID signals.} In previous work \cite{ChenYou15,HutchBea18}, rhythmic taps are directly performed on mobile devices and are fairly easy to detect from inertial sensors. In contrast, rhythmic taps in RF-Rhythm are performed on the RFID card and have to be indirectly extracted from noisy backscattered signals. We explore various signal processing techniques to process noisy raw phase data for extracting a reliable tapping rhythm. We also use machine learning techniques to train a classifier the backend server uses to validate an extracted tapping rhythm.

\textbf{The second challenge is anti-eavesdropping, i.e., how to prevent the adversary from acquiring the user's tapping rhythm from sniffed RFID signals.} In particular, the adversary can easily eavesdrop on the open RFID channel and then behave in the same way as the RFID reader to decode the user's tapping rhythm from intercepted RFID signals. It can then repeat the rhythmic taps on lost/stolen/cloned RFID card to successfully impersonate the legitimate user. We tackle this challenge by a novel phase-hopping protocol in which the RFID reader emits Continuous Wave (CW) with random phases for extracting the user's tapping rhythm. Since the adversary does not know the phase-hopping sequence, it can no longer extract the correct tapping rhythm from sniffed RFID signals. 

We thoroughly evaluate the security and usability of RF-Rhythm by comprehensive experiments on Impinj RFID readers, COTS passive tags, and USRP devices. Our experiments involve 19 volunteers from two countries and explore three representative machine learning techniques, including  Support Vector Machine (SVM), Neural Networks (NN), and Convolutional Neural Networks (CNN). We show that RF-Rhythm is highly secure with false-positive and false-negative rates close to zero. In addition, we demonstrate the high resilience of RF-Rhythm to brute force, visual eavesdropping, and RF eavesdropping attacks. We also confirm the high usability of RF-Rhythm by a user survey. 

The rest of this paper is organized as follows. Section~\ref{sec:background} gives some necessary background about RFID systems. Section~\ref{sec:attack_models} describes the adversary model. Section~\ref{sec:overview} provides an overview of RF-Rhythm. Section~\ref{sec:RF-FHYTHM-Design} details the design of RF-Rhythm. Section~\ref{sec:anti-eaves} presents the phase-hopping protocol for anti-eavesdropping. Section~\ref{sec:evaluation} reports the experimental evaluation of RF-Rhythm. Section~\ref{sec:related} briefs the related work.

\section{Basics of Passive UHF RFID Systems}\label{sec:background}
In this section, we introduce some necessary background about passive Ultra-High-Frequency (UHF) RFID systems to help illustrate the RF-Rhythm design later. An RFID system consists of a backend server, readers, and RFID cards. The RFID reader sends both modulated commands and continuous wave (CW). The RFID card sends back its data by exploring the energy harvested from the reader's signals to switch its input impedance between two states and thus modulate the backscattered signal. EPC Gen 2 \cite{UHFG218} is the most popular UHF RFID standard and assumed throughout the remainder of this paper.

%\subsection{FM0 Baseband}

\begin{figure}[t]
	\centering
	\includegraphics[width=3.2in]{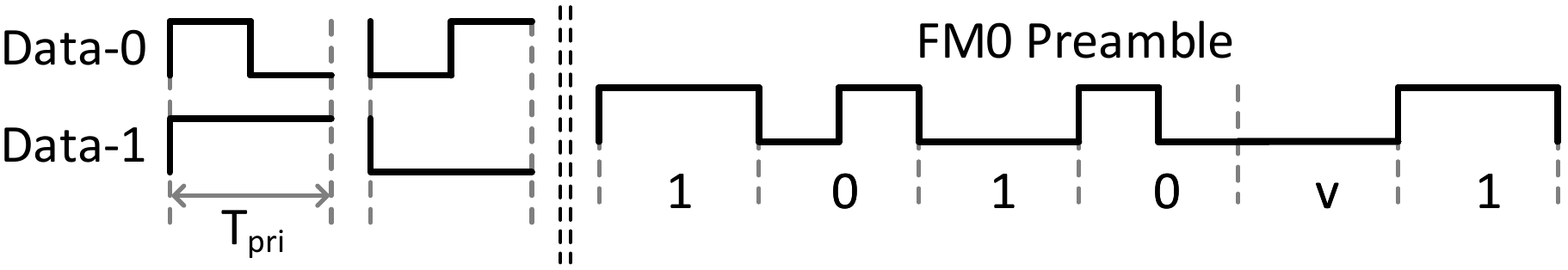}
	\caption{FM0 baseband symbols and preamble.}
	\label{fig:fm0_baseband}
	\vspace{-.15in}
\end{figure}

\begin{figure}[h]
	\centering
	\includegraphics[width=3.4in]{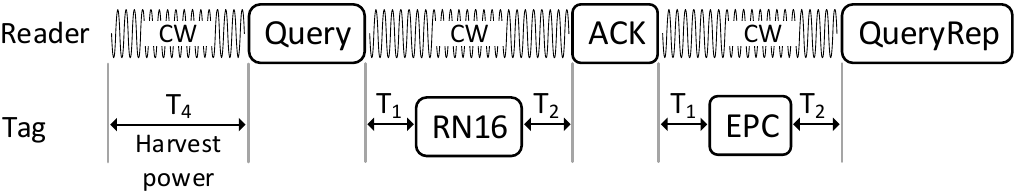}
	\caption{The basic EPC Gen-2 query protocol with a single RFID card.}
	\label{fig:protocol_overview}
\end{figure}

RFID cards encode the backscattered data using either FM0 baseband or miller modulation. We only consider FM0 encoding in this paper, but our work can easily extend to miller modulation. Fig.~\ref{fig:fm0_baseband} shows the basic FM0 symbols. FM0 inverts the baseband phase at every symbol boundary with an additional mid-symbol phase inversion for each data-0. The duration of an FM0 symbol is denoted by $T_\textrm{pri}=1/\textrm{BLF}$, where BLF represents the backscatter link frequency ranging from $40$~kHz to $640$~kHz \cite{UHFG218}. To ease our presentation, we assume BLF equal to 40~kHz, corresponding to $T_\textrm{pri}=25\mics$.

Fig.~\ref{fig:protocol_overview} shows the basic query protocol in EPC Gen-2 \cite{UHFG218}. 
\begin{enumerate}
	\item The reader emits CW of length $T_4$ for the RFID card to harvest and store energy.
	\item The reader sends a Query command followed by CW of length $T_{1} + T_{2} + T_{\rm{RN16}}$. During this CW period, the card backscatters an RN16 message comprising a 6-bit preamble, a 16-bit random number, and one dummy bit. 
	\item The reader sends an ACK followed by CW of length $T_{1} + T_{2} + T_{\rm{EPC}}$. During this CW period, the card backscatters its EPC (Electronic Product Code). 
	\item The reader sends QueryRep to finish this query session. 
\end{enumerate}

EPC Gen-2 \cite{UHFG218} gives recommendations for the above timing parameters. Let RTcal represent the duration of Interrogator-to-Tag calibration symbol, which is specified in the reader configuration and set to $\rm RTcal = 72\mics$ in our implementation. Also let $\rm FrT$ be the frequency tolerance of FM0 baseband signals, which equals 4\% for $\textrm{BLF = 40 KHz}$. We have $T_4=\textrm{2RTcal}=144\mics$ and $75\mics \leq T_{2} \leq 500\mics$. In addition, the maximum, minimum, and nominal values of $T_{1}$ are $262\mics$, $238\mics$,  and $250\mics$, respectively.

\section{Adversary Model}\label{sec:attack_models}

We assume an adversary $\mathcal{A}$ who attempts to use a lost/stolen/cloned RFID card to pass authentication checks and thus impersonate the legitimate card user. $\mathcal{A}$ knows how RF-Rhythm works and can perform rhythmic taps on the RFID card with fingers or even a fully programmable robotic arm. We assume that $\mathcal{A}$ does not know the legitimate user's secret song segment and can try the following attack strategies.  
\begin{itemize}
	\item \textbf{Brute force}: $\mathcal{A}$ performs totally random rhythmic taps. 
	
	\item \textbf{Visual eavesdropping}: $\mathcal{A}$ observes the legitimate user's tapping behavior, e.g., by shoulder surfing or a spy camera, and then tries to emulate it. 
	
	\item \textbf{RF eavesdropping}: $\mathcal{A}$ sniffs all the PHY communication traces between the RFID reader and card to recover and then perform the legitimate user's rhythmic taps. 
\end{itemize} 

\begin{figure}[t]
	\centering
	\includegraphics[width=3.4in]{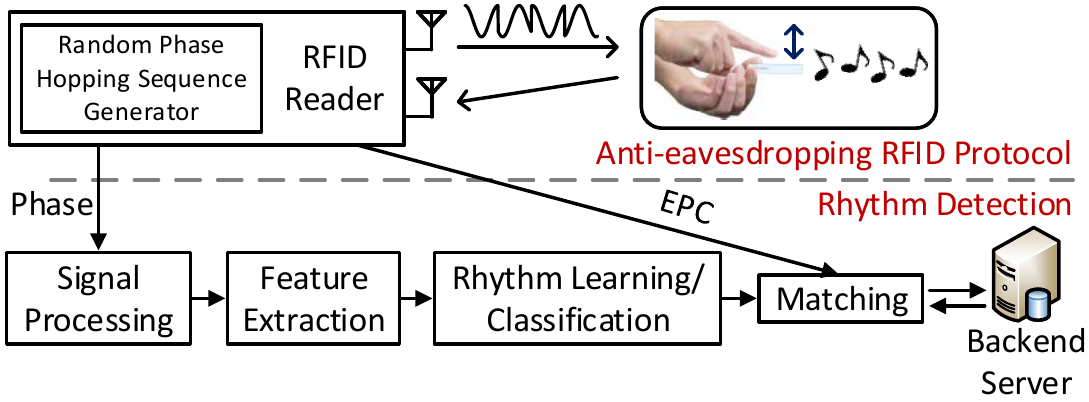}
	\caption{The RF-Rhythm system flowchart.}
	\label{fig:overview}
	\vspace{-.15in}
\end{figure}

\section{System Overview}\label{sec:overview}

RF-Rhythm consists of an enrollment phase and a verification phase, and its major modules are depicted in Fig.~\ref{fig:overview},  

During the enrollment phase, the legitimate user first selects an arbitrary song segment familiar to him/herself. Then the user performs rhythmic taps on his/her RFID card in accordance with his/her own interpretation of the chosen song segment, e.g., by singing it silently. The user's tapping rhythm is referred to as his/her secret rhythm hereafter. 

The security of RF-Rhythm relies on the secrecy of the chosen song segment and also the user's likely unique tapping rhythm. In particular, since there are numerous songs available, the adversary can hardly guess the selected song segment of a target user; an advanced user such as a musician can even self-compose the song segment. In addition, people may have very subjective mental interpretations about the same song segment, resulting in totally different tapping rhythms. 

The backend server handles the enrollment request as follows. First, it acquires the EPC of the user's RFID card through the reader as usual by using the protocol in Fig.~\ref{fig:protocol_overview}. Second, it instructs the user to perform rhythmic taps on the RFID card, which would lead to phase changes in the backscattered signals received by the reader. Third, the server invokes a \emph{Signal Processing} module to extract reliable phase data from noisy backscattered signals. Fourth, it uses a \emph{Feature Extraction} module to obtain a feature vector that characterizes the use's tapping rhythm. Finally, it asks the user to repeat the rhythmic taps multiple times and then feeds all the resulting feature vectors into a \emph{Rhythm Learning} module to train a high-quality binary rhythm classifier for this user. 

In the verification phase, the backend server first explores the RFID card for its EPC with the protocol in Fig.~\ref{fig:protocol_overview}. If the EPC is found in the database, the server instructs the reader to execute multiple rounds of the protocol again in Fig.~\ref{fig:protocol_overview}. RF-Rhythm is highly usable in the sense that the RFID user just needs to perform his/her secret tapping rhythm multiple times without the need to know when the server starts to extract it in both the enrollment and verification phases. The server invokes the same Signal Processing and Feature Extraction modules to extract a candidate tapping rhythm in each round, which is then tested with the trained rhythm classifier associated with the EPC acquired before. The authentication process terminates until when the server either detects a valid tapping rhythm or fails to detect one after a threshold number of rounds. The RFID card and corresponding user are considered authentic in the former case and fake in the latter.  

RF-Rhythm features a novel anti-eavesdropping protocol employed by the RFID reader to emit CW with random phases for extracting the user's secret tapping rhythm in both enrollment and verification phases. Our protocol can prevent a capable adversary from recovering and then replaying the legitimate user's secret rhythm from sniffed RFID signals. 

Our descriptions above focus on very cheap COTS RFID cards and can also be easily adapted to more powerful, expensive cryptographic RFID cards. For example, the EPC can just be replaced by a cryptographic authentication message. We ignore this option henceforth for ease of illustration.

\section{RF-Rhythm Design Details}\label{sec:RF-FHYTHM-Design}

\subsection{Feasibility Study: Tap Detection}\label{sec:ext_phase}

The backscattered signal's phase is readily available on commercial RFID readers such as Impinj R420 \cite{RX420}. According to \cite{LLUD13}, it can be expressed as $\phi = (\frac{4\pi d f}{c}+\phi_{\textrm{reader}}+\phi_{\textrm{card}}) \mod 2\pi$, where $2d$ is the round-trip propagation distance between the reader and card, $f$ is the CW frequency, $c$ is the speed of light, $\phi_{\textrm{reader}}$ denotes the phase rotation due to the reader's transmit and receive circuits,  and $\phi_{\textrm{card}}$ represents the phase rotation caused by the RFID card's reflection characteristics. 

Finger taps on the RFID card can change its circuit impedance and also signal propagation, leading to some additional phase rotation denoted by $\phi_{\textrm{tap}}$. So we modify the phase expression above to 
\begin{equation}\label{eq:phaseTap}
\phi = \Big(\frac{4\pi d f}{c}+\phi_{\textrm{reader}}+\phi_{\textrm{card}}\Big)+ \phi_{\textrm{tap}} \mod 2\pi. 
\end{equation}

\begin{figure}[t]
	\centering
	\includegraphics[width=3.2in]{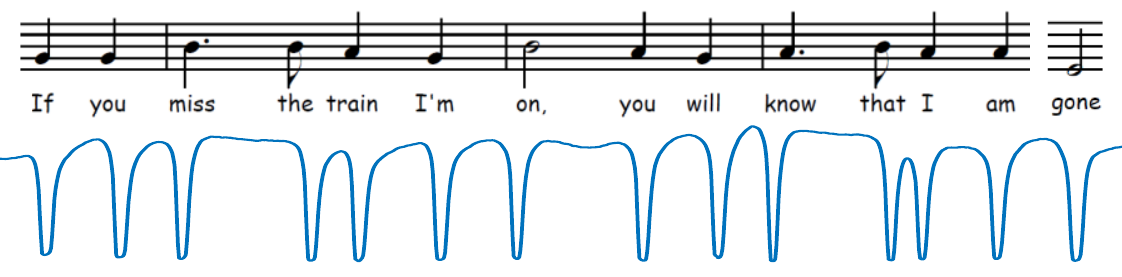}
	\caption{Absolute phase changes induced by rhythmic taps on an RFID card.}
	\label{fig:cp_phase_rssi}
	\vspace{-.15in}
\end{figure}

\begin{figure}[t]
	\centering
	\subfloat[Absolute phase]{
		\centering
		\label{fig:touch_phase}
		\includegraphics[width=1.6in]{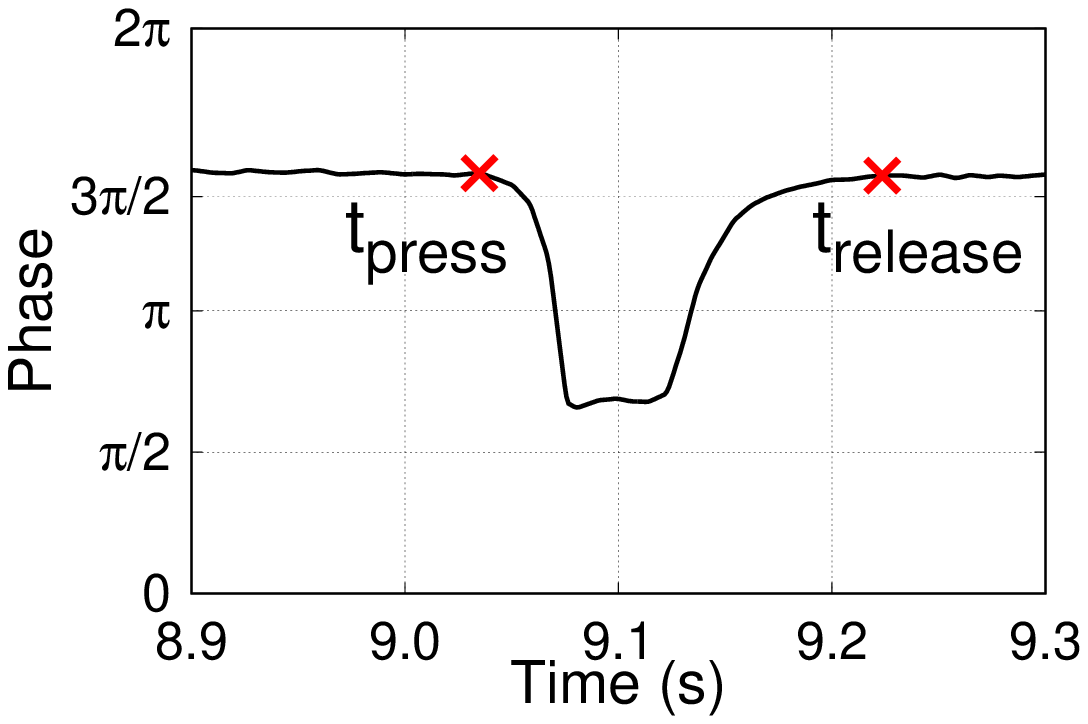}}
	\hfill
	\centering
	\subfloat[Phase difference]{
		\centering
		\label{fig:touch_delta_phase}
		\includegraphics[width=1.6in]{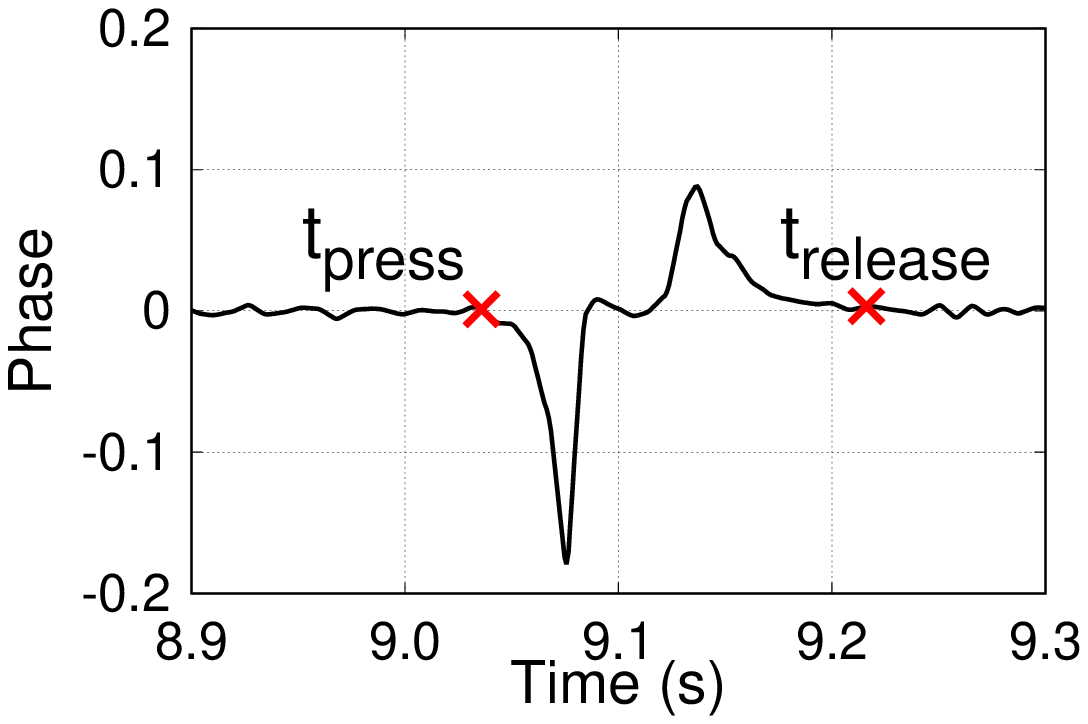}} % using un-filtered data to plot the delta phase
	\caption{Absolute and differential phase changes caused by a single tap.}\label{fig:phaseOneTap}
	\vspace{-.15in}
\end{figure}

To better understand the effect of finger taps, we perform a simple experiment using a Impinj R420 reader and a SMARTRAC R6 DogBone tag. Fig.~\ref{fig:cp_phase_rssi} shows the phase changes induced by rhythmic finger taps on the RFID card in accordance with the shown song segment. We also show the phase change associated with a single tap in Fig.~\ref{fig:phaseOneTap}. A tap event can be decomposed into a press stage and a release stage. So we use $[t_{\textrm{press}}, t_{\textrm{release}}] $ to represent a tap event in the time domain, where $t_{\textrm{press}}$ and $t_{\textrm{release}}$ denote the time that the phase (difference) starts to change and return to the baseline value, respectively. Fig.~\ref{fig:touch_phase} and Fig.~\ref{fig:touch_delta_phase} depict the absolute phase values and the difference between adjacent phase values, respectively. These results clearly demonstrate the feasibility of exploring phase changes for tap detection. 

\subsection{Data Processing}

We represent the reader's phase report at time $t_i$ by $[\phi_{i}, f_{i}, t_{i}]$, where $f_i$ denotes the CW frequency at $t_i$. According to Eq.~(\ref{eq:phaseTap}), we have 
\begin{equation}\label{eq:phaseTi}
\phi_i = \Big(\frac{4\pi d f_i}{c}+\phi_{\textrm{reader}}+\phi_{\textrm{card}}\Big) + \phi_{\textrm{tap}, i} \mod 2\pi\;, 
\end{equation}
where $\phi_{\textrm{tap}, i}$ denotes the phase shift during the $i$th tap. The interval $t_{i+1}-t_i$ ($i\geq 0$) is about $4\ms$ on the Impinj R420 reader. We temporarily assume that $f_i$ is constant and perform the following steps to process the raw phase data to extract more useful information for further rhythm extraction. 

\vspace{.03in} \noindent \textbf{Phase difference and unwrapping.} We use the phase difference instead of the absolute phase to eliminate the approximately constant  $\frac{4\pi d f_i}{c}+\phi_{\textrm{reader}}+\phi_{\textrm{card}}$ during adjacent tap events. In addition, the raw phase data are wrapped within $[0, 2\pi]$, so it is critical to perform phase unwrapping to eliminate ambiguity. Our experiments reveal that although the phase change induced by tap events are sharp, it is always bounded by $\pi$. According to this finding, the unwrapped phase difference is calculated by
\begin{equation}\label{eq:deltaPhase}
\Delta\phi_{i} = \phi_{\textrm{tap}, i}-\phi_{\textrm{tap}, i-1}= \left\{\begin{array}{ll}
\phi_{i} - \phi_{i - 1}, & \abs{\phi_{i} - \phi_{i - 1}} \leq \eta \\
\phi_{i} - \phi_{i - 1} + 2\pi, & \phi_{i} - \phi_{i - 1} < -\eta\\
\phi_{i} - \phi_{i - 1} - 2\pi, & \phi_{i} - \phi_{i - 1} > \eta
\end{array}\right.
\end{equation}
Here $\eta$ is an empirical value set to 3.5 in this paper.

\vspace{.03in}\noindent \textbf{Normalization.} Since the sampling rate of the RFID reader is not consistent, so we further derive the time-normalized phase difference as 
\begin{equation}
\overline{\Delta\phi_{i}} = \frac{\Delta\phi_{i}}{\Delta t_{i}}=\frac{\Delta\phi_{i}}{t_{i} - t_{i - 1}}\;.
\end{equation}

\vspace{.03in}\noindent \textbf{Interpolation and filtering.} We further use a linear interpolation with a factor of 4 and a 15-point average value filter to smooth the data and also mitigate the noise. We denote the final smoothed data by $\Phi = [\overline{\Delta\phi_{1}}, \overline{\Delta\phi_{2}}, \dots,\overline{\Delta\phi_{N}}]$, where $N$ denotes the total number of data points. 

\subsection{Mitigating Frequency Hopping}\label{sec:mitigateFHSS}

\begin{figure}[h]
	\centering
	\subfloat[Raw phase with frequency hopping]{
		\centering
		\label{fig:original_phase}
		\includegraphics[height=1in]{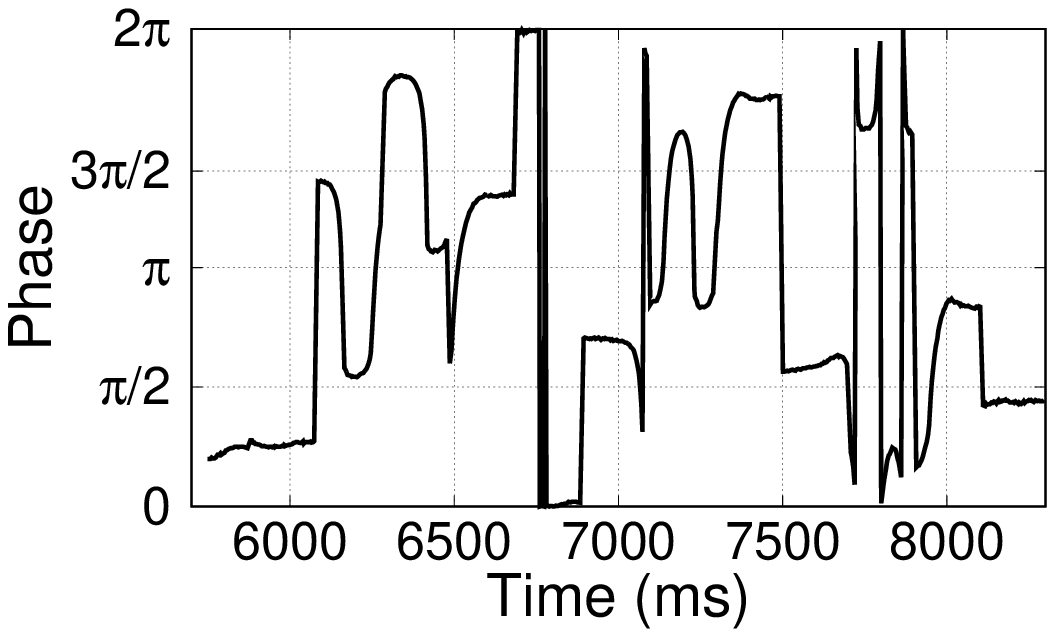}}
	\hfill
	\centering
	\subfloat[Processed phase difference]{
		\centering
		\label{fig:processed_phase}
		\includegraphics[height=1in]{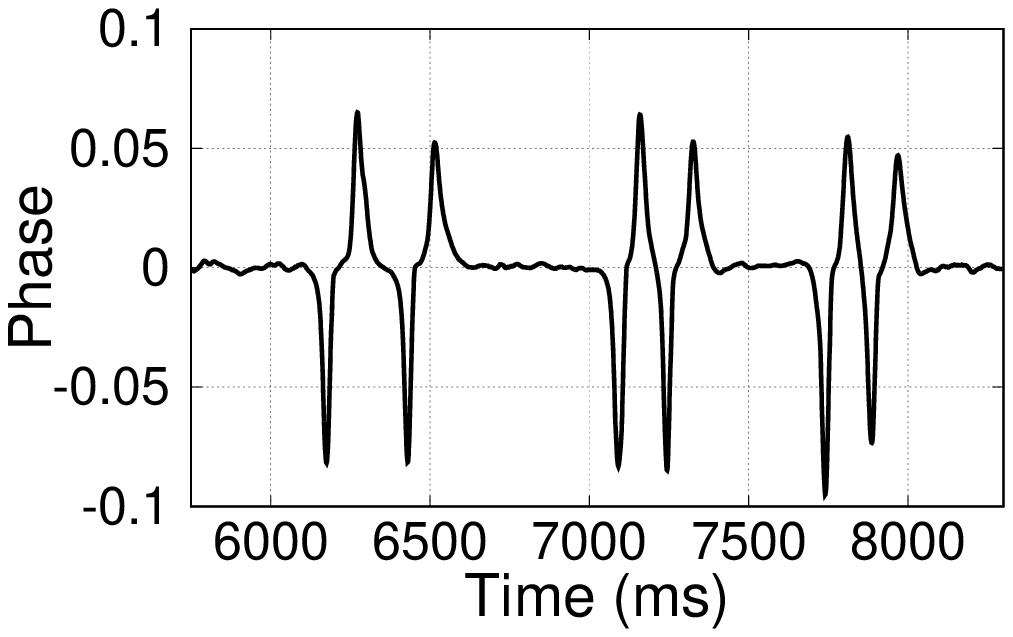}} % using un-filtered data to plot the delta phase
	\caption{Data processing under frequency hopping.}
	\label{fig:freq_hopping}
	\vspace{-.15in}
\end{figure}

We intend RF-Rhythm to be a universal solution worldwide and thus must deal with frequency hopping mandated in many regions. For example, FCC requires that all RFID readers used in the US apply frequency hopping across 50 channels ranging from 902 to 928 MHz with the dwell time on each interval no larger than 0.4 seconds. According to Eq.~(\ref{eq:phaseTi}), such frequency hopping naturally leads to phase discontinuity in Fig.~\ref{fig:original_phase}.

To see the effect of frequency hopping more clearly, assume that frequency hopping occurs at $t_i$ ($i\geq 2$). In the Impinj R420 reader, the frequency-hopping interval is $200\ms$, while the phase-report interval is about $4\ms$. So there is no frequency hopping at $t_{i-2}$, $t_{i-1}$, and $t_{i+1}$, i.e., $f_{i-2}=f_{i-1}\neq f_i=f_{i+1}$. The phase difference in Eq.~(\ref{eq:deltaPhase}) is in effect 
\[
\Delta\phi_{i} = \phi_{\textrm{tap}, i}-\phi_{\textrm{tap}, i-1}+\big(\frac{4\pi d f_{i}}{c} - \frac{4\pi d f_{i - 1}}{c}\big). 
\]
Since $d$ is unknown and hard to estimate in practice, we cannot do a simple calibration by subtracting the term in the parenthesis from $\Delta\phi_{i}$. Instead, we compute the time-normalized phase difference for $t_i$ as 
\begin{equation}
\overline{\Delta\phi_{i}} = (\overline{\Delta\phi_{i + 1}} + \overline{\Delta\phi_{i - 1}})\frac{t_{i} - t_{i - 1}}{t_{i + 1} - t_{i - 1}}
\end{equation}
Fig.~\ref{fig:processed_phase} plots the output of the Data Processing module corresponding to Fig.~\ref{fig:original_phase} after we adopt the above technique. 

\subsection{Feature Extraction}

Since a tapping rhythm consists of individual taps and tap-durations, we first seek to extract individual tap events from the processed phase data $\Phi = [\overline{\Delta\phi_{1}}, \overline{\Delta\phi_{2}}, \dots,\overline{\Delta\phi_{N}}]$. Recall that each tap event can be represented by $[t_{\textrm{press}}, t_{\textrm{release}}]$. We draw three observations from Fig.~\ref{fig:touch_delta_phase} obtained from preliminary experiments. First, the start and end of a tap event correspond to the phase difference beginning to deviate from and return to the zero baseline, respectively. Second, the phase difference first decreases from and then returns to the zero baseline when the user finger goes from just touching to fully pressing on the RFID card, leading to a local minimum. Finally, the phase difference first increases from and then returns to the zero baseline when the user finger goes from decreasing the pressure on to completely leaving the RFID card, resulting in a local maximum. The later two observations are both because the card impedance gradually change with the finger pressure on the card during a tap event.  Armed with these observations, we use the following empirical process

\begin{enumerate}
	\item Find all the local maximums above $\delta_1$ and minimums below $\delta_2$ in $\Phi$.
	\item Pair each local minimum with the immediate local maximum (if any) such that there are no other local minimums or maximums in between. We require the user's tapping rhythm to be sufficiently long such that $M\gg 2$ local minimum-maximum pairs can be located in $\Phi$, each associated with a unique tap event.	
	\item Find the first data point before (after) the local minimum (maximum) which is within $\pm \delta_3$ from the zero baseline for each local minim-maximum pair. The corresponding timestamp is used as $t_{\textrm{press}}$ ($t_{\textrm{release}}$) of the tap event. 	
\end{enumerate}
The thresholds $\delta_1$, $\delta_2$, and $\delta_3$ can be obtained empirically through experiments. 

Finally, we obtain an $M$-tap event sequence as
\begin{equation}
\mathbb{V} = \left[
\begin{array}{cccc}
t_{\textrm{press}, 1} 	& t_{\textrm{press}, 2} 	  & \dots & t_{\textrm{press}, M}\\
t_{\textrm{release}, 1}  & t_{\textrm{release}, 2}  & \dots & t_{\textrm{release}, M}\\
\end{array}\right]\;,
\end{equation}
from which we can derive a feature vector $\mathbb{F}=[F_1,\dots,F_{M-1}]$, where $F_i=t_{\textrm{press}, i+1} - t_{\textrm{release}, i}$. 

\subsection{Rhythm Classification}

The backend server builds a rhythm classifier during the enrollment phase. To do so, it instructs the user to perform rhythmic taps in accordance with his/her secret song segment multiple times. The resulting phase-difference vectors may vary due to slight tapping variations. So we apply Dynamic Time Warping (DTW) \cite{VemulHum14} to align all the phase-difference vectors to that of the first acquired tapping rhythm. Then we obtain a feature vector from each aligned phase-difference vector and pad zeros in the end (if needed) to make all the feature vectors have the same length. Finally, we use the resulting feature vectors to train a rhythm classifier based an any established machine learning technique. We compare the performance of one-vs-all linear Support Vector Machine (SVM), Neural Networks (NN), and Convolutional Neural Networks (CNN) in Section~\ref{sec:evaluation}. During each authentication session, the server explores the same processes to extract a tapping rhythm and then test it with the rhythm classifier.

\section{Anti-Eavesdropping via Phase Hopping}\label{sec:anti-eaves}

In this section, we present a novel phase-hopping technique to prevent a capable adversary from acquiring the legitimate tapping rhythm from sniffed RFID signals. In what follows, we first illustrate the rhythm-eavesdropping attack, followed by the motivation for using phase hopping as a defense. Then we detail the protocol design and analyze its security.

\begin{figure}[t]
	\centering
	\includegraphics[height=1.0in]{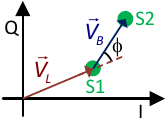}
	\caption{Complex demodulated signals received by the reader.}
	\label{fig:ext_phase}
	\vspace{-.15in}
\end{figure}

\subsection{Rhythm-Eavesdropping Attack}\label{sec:RhythmEave}

We first explain the principle with which the RFID reader extracts the signals backscattered by the RFID card. As shown in Fig.~\ref{fig:fm0_baseband}, there are two possible voltage levels in FM0 symbols. The card only backscatters when transmitting high-voltage pulses. Consider the query protocol in Fig.~\ref{fig:protocol_overview}. The symbols received by the reader between its two consecutive commands (e.g., Query and ACK) can be classified into two states (S1 and S2). The symbols in S1 contain only constant CW, while those in S2 are the superposition of CW and backscattered signals. For simplicity, we represent the symbols in S1 and S2 by two single points in the complex I-Q plane in Fig.~\ref{fig:ext_phase}, corresponding to vector $\vec{V_{L}}$ and $\vec{V_{B}}$, respectively. The phase of backscattered signals can be derived as \cite{WangMov16}
\begin{equation}\label{eq:ext_phase}
\phi = \arccos(\frac{\vec{V_{B}} \cdot \vec{V_{L}}}{\abs{\vec{V_{B}}} \abs{\vec{V_{B}}} }).
\end{equation}
The phase reports from the reader correspond to the samples of $\phi$ above. As said, the phase-sampling frequency in the Impinj R420 reader is about $4\ms$. 

To launch the rhythm-eavesdropping attack, the adversary can just passively sniff the reader-card communications with its own RFID reader or a software-defined radio. After classifying sniffed symbols into S1 and S2, it uses the same process above to extract $\phi$. Next, it explores the workflow in Section~\ref{sec:RF-FHYTHM-Design} to acquire the legitimate tapping rhythm. Finally, it can carefully study the tapping rhythm and reproduce it by hand or even through a programmable robotic arm on the lost/stolen/cloned RFID card. Since this attack directly exploits physical-layer RFID signals, it cannot be thwarted by encrypting RFID protocol messages at the application layer. 
\begin{figure}[t]
	\centering
	\subfloat[]{
		\centering
		\label{fig:jam_phase_wave}
		\includegraphics[height=0.9in]{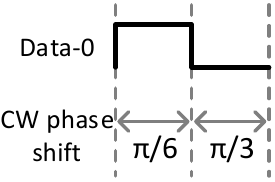}}
	\hfill
	\centering
	\subfloat[]{
		\centering
		\label{fig:jam_phase}
		\includegraphics[height=1.1in]{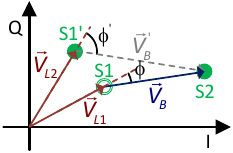}} % using un-filtered data to plot the delta phase
	\caption{Illustration of reader-phase hopping.}\label{fig:phase-hopping}
	\vspace{-.15in}
\end{figure}

\subsection{Phase Hopping to Mitigate Rhythm Eavesdropping} \label{sec:change_phase_analysis}

We propose to let the RFID reader emit CW with random phases to counteract the rhythm-eavesdropping attack. The objective is to prevent the adversary from obtaining matching symbols in states S1 and S2, so it cannot derive the correct phases of backscattered signals as in Fig.~\ref{fig:ext_phase}. 

Fig.~\ref{fig:phase-hopping} explains the intuition of our defense. Assume that the RFID card is backscattering a data-0 symbol. As said above, the card only backscatters the high-voltage part. As shown in Fig.~\ref{fig:jam_phase_wave}, we let the reader set the CW phases to $\pi/6$ and $\pi/3$ during backscattering and non-backscattering, respectively. The adversary again tries to cluster sniffed symbols into states S1 and S2. Due to phase hopping, the S1 symbols that correspond to non-backscattering has a phase offset of $\pi/3$, labeled by $\textrm{S1}'$ in Fig.~\ref{fig:jam_phase}. The true S1 symbol matching the S2 symbol, however, should have a phase offset of $\pi/6$, labeled by S1 in Fig.~\ref{fig:jam_phase}. Since the adversary does not know the true CW phase during backscattering, it can only use the symbols in $\textrm{S1}'$ and S2 to derive a wrong phase $\phi'$. But the reader knows the true CW phase or S1 symbol and can thus derive the correct phase $\phi$. 

\subsection{Protocol Design}

It is very challenging to properly implement the phase-hopping idea above. In particular, our example in Fig.~\ref{fig:phase-hopping} assumes perfect reader-tag synchronization such that the reader knows exactly when backscattering occurs and thus when to change the CW phase. This assumption is impossible to hold in practice. Therefore, the adversary may still be able to obtain matching symbols in S1 and S2 to derive the correct phase and eventually the legitimate tapping rhythm. A tempting solution is using a very short hopping interval, which nevertheless may negatively affect the reader's capability to recover the correct phase and thus the tapping rhythm. It is thus critical to determine the optimal phase-hopping interval to strike a balance between attack resilience and system correctness.   

We illustrate our phase-hopping protocol with a simplified version of the query protocol in Fig.~\ref{fig:protocol_overview}. Assume that the backend server acquires and validates the card's EPC with the protocol in Fig.~\ref{fig:protocol_overview}. It then instructs the RFID reader to initiate additional query rounds to acquire the user's tapping rhythm. Each query round consists of a Query message followed by a CW period of length $T_{1} + T_{2} + T_{\rm{RN16}}$, where $T_1$ and $T_2$ are random variables mentioned in Section~\ref{sec:background}. In the original RFID protocol, the CW phase is constant. Our goal now is to determine when phase hopping should start/stop and how often it should be in each CW period. 

The begin and end of phase hopping depend on $T_1$. According to Section~\ref{sec:background}, $T_{1}$ is in $[238\mics, 262\mics]$ with the nominal value equal to  $250\mics$. We also measure the actual distribution of $T_{1}$ over 5,639 card replies. Since 98.92\% of $T_1$ are between $244\mics$ to $247\mics$, it is safe to conclude that if the phase-hopping duration covers $[244\mics, 247\mics + T_{\textrm{RN16}}]$, almost all the backscattered signals associated with RN16 can be covered. 

\begin{figure}[t]
	\centering
	\includegraphics[width=3.4in]{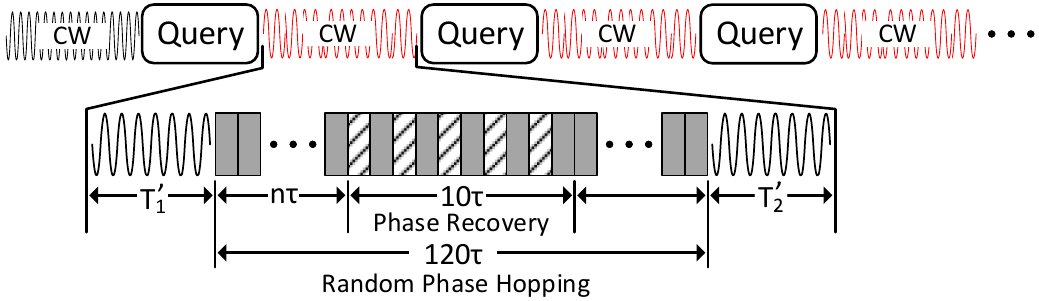}
	\caption{Timing diagram of phase hopping.}
	\label{fig:jam_protocol}
	\vspace{-.15in}
\end{figure}

\begin{figure*}[t]
	\centering
	\subfloat[Received signals during the phase-hopping duration.]{
		\centering
		\label{fig:jam_sig_tag}
		\includegraphics[height=1.1in]{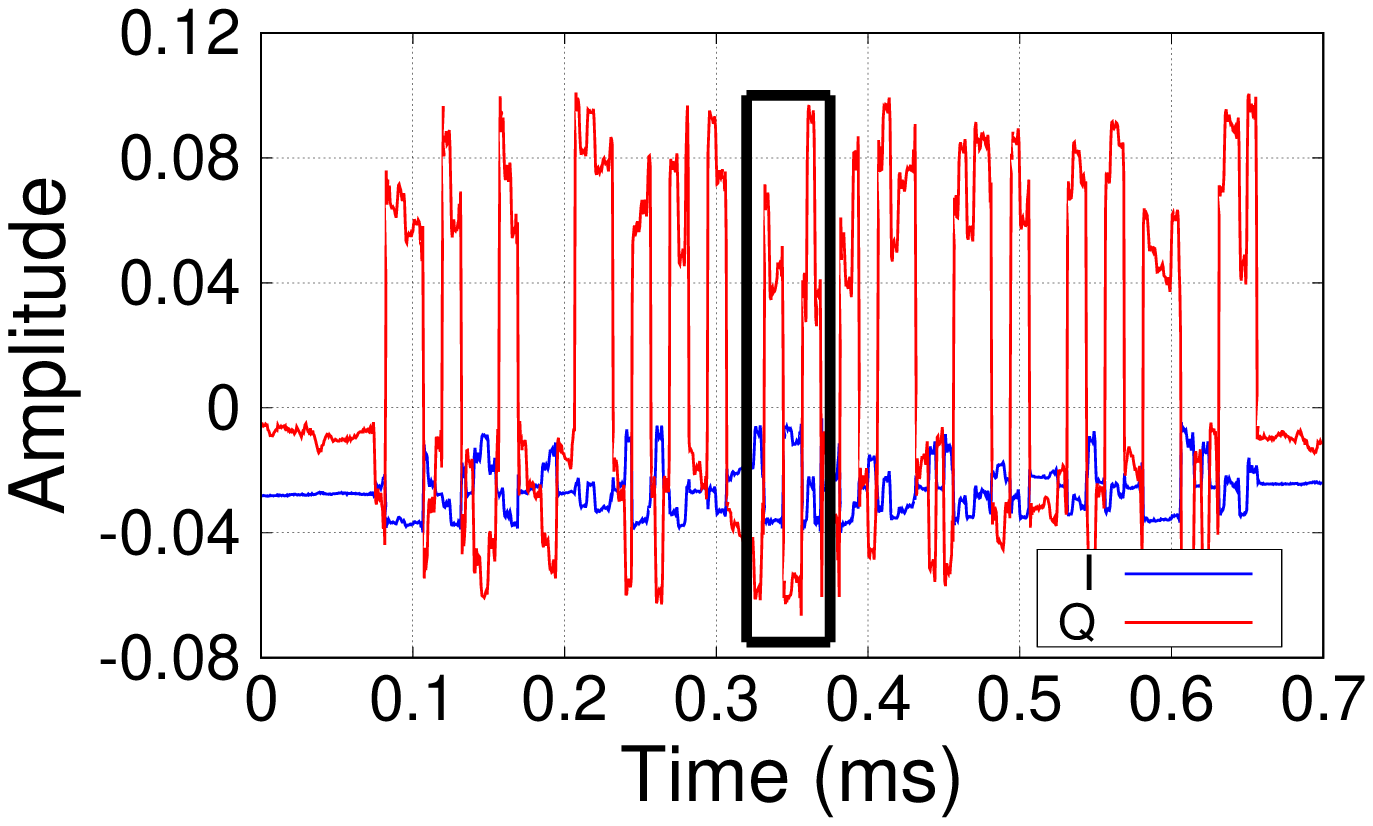}}
	\hfill
	\centering
	\subfloat[Extracted symbols for $\theta_{\textrm{recover}}$ in the phase-recovery period. ]{
		\centering
		\label{fig:jam_sig_tag_const}
		\includegraphics[height=1.1in]{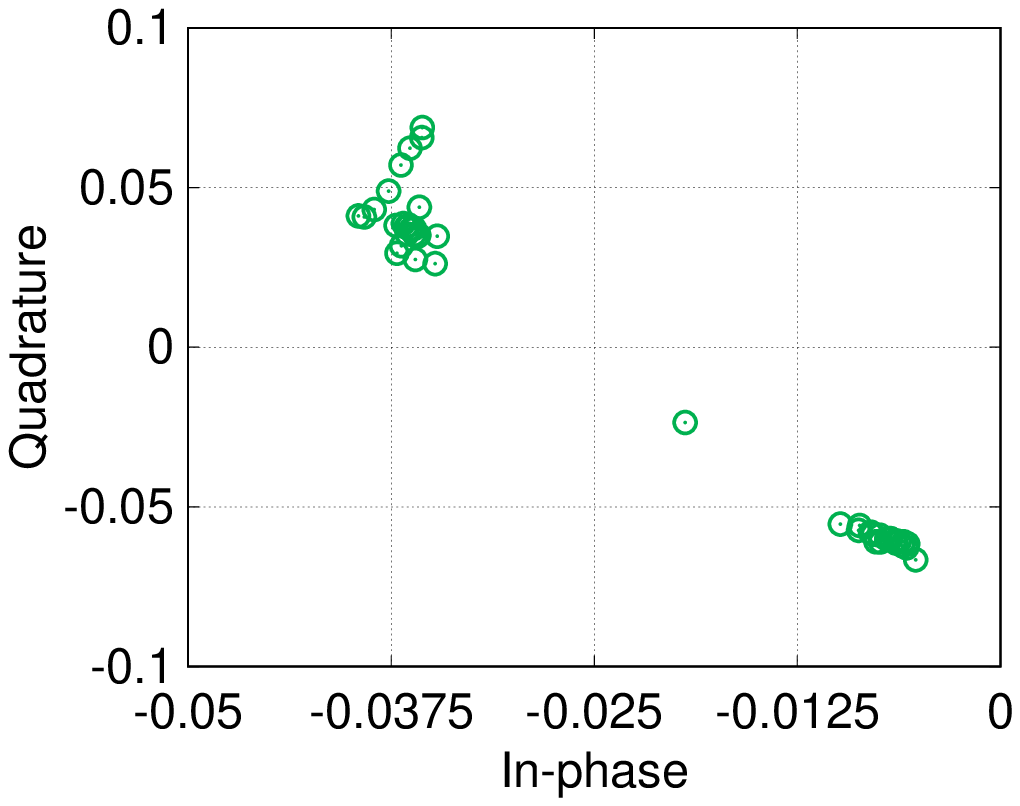}}
	\hfill
	\centering
	\subfloat[The adversary's sniffed symbols for all signals in Fig.~\ref{fig:jam_sig_tag}.]{
		\centering
		\label{fig:jam_sig_tag_const_all}
		\includegraphics[height=1.1in]{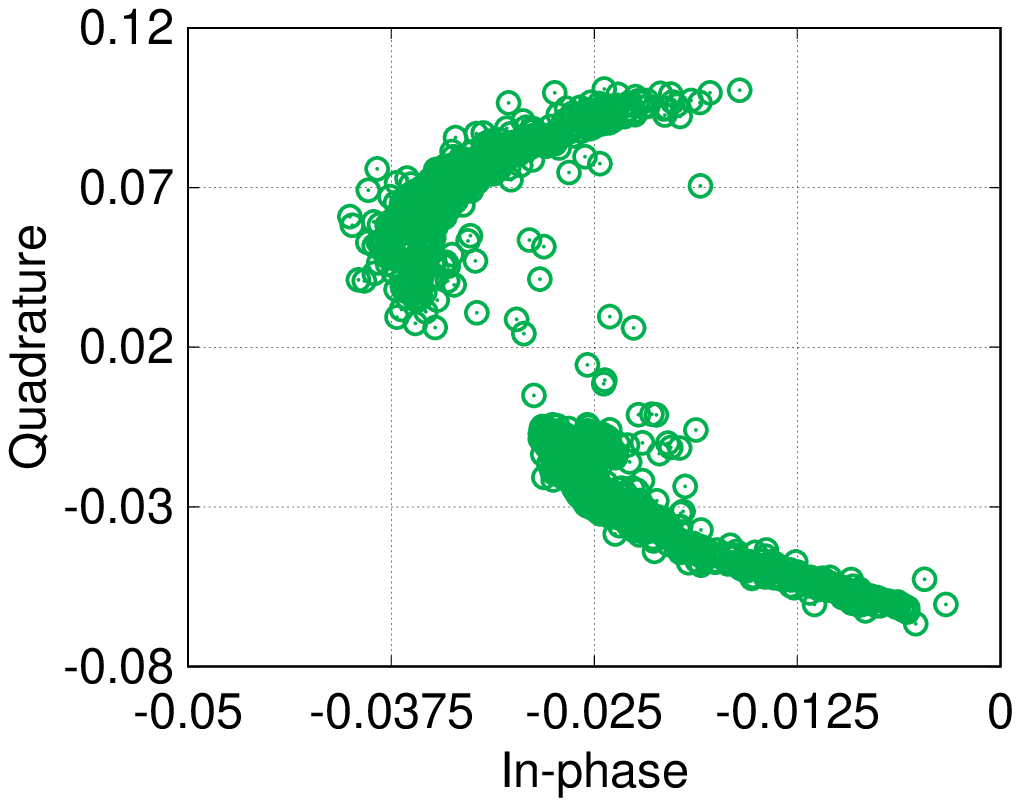}}
	\hfill
	\centering
	\subfloat[Phase recovered by the reader.]{
		\centering
		\label{fig:dist_phase}
		\includegraphics[height=1.1in]{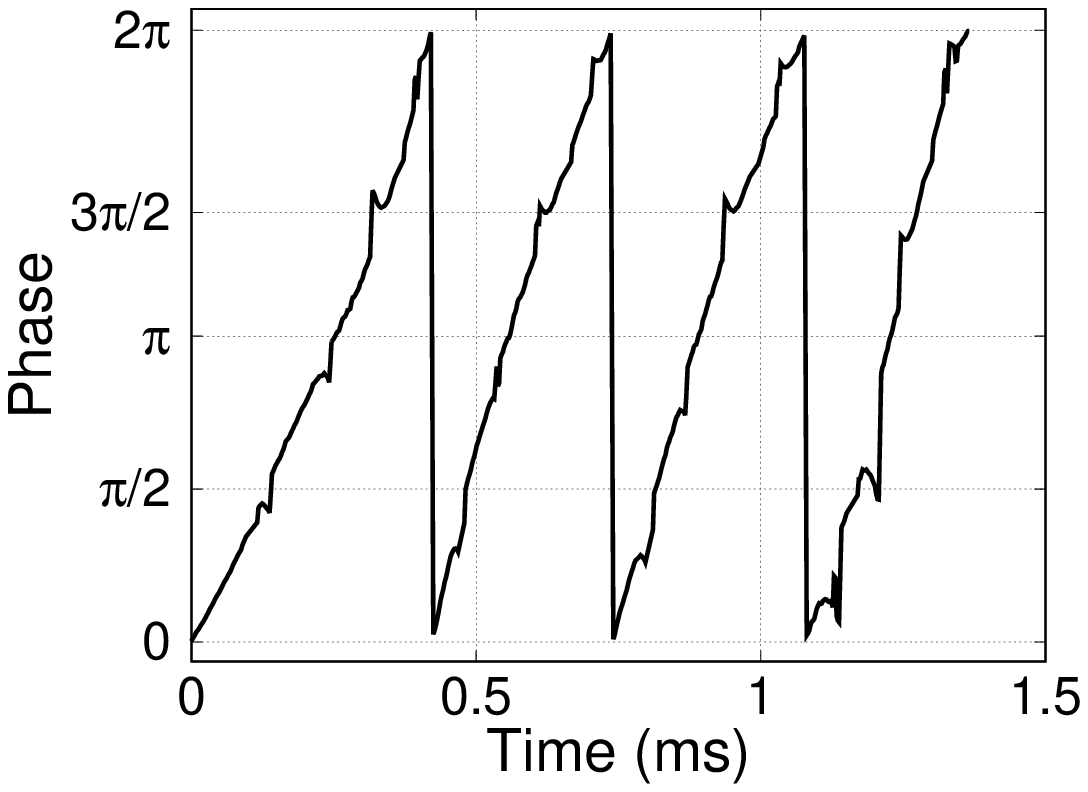}}
	\caption{Effect of reader-phase hopping.}
	\label{fig:hoppingEffect}
	\vspace{-.15in}
\end{figure*}

The next challenge is to determine the hopping interval $\tau$, which should be as short as possible for high attack resilience. The minimum $\tau$ is hardware-specific and empirically set to $\tau=\frac{T_\textrm{pri}}{5}=5\mics$ in our USRP implementation, where  $T_\textrm{pri}=1/\textrm{BLF}=25\mics$ denotes the FM0 symbol duration introduced in Section~\ref{sec:background}. Ideally speaking, each CW phase value leads to a unique pair of S1 and S2 symbols as shown in Fig.~\ref{fig:ext_phase}. In practice, we can only obtain two clusters of symbols associated with S1 and S2, respectively, which are referred to the S1 and S2 clusters for convenience. The RFID reader needs to obtain the matching S1 and S2 clusters for at least one random CW phase to recover the correct phase for the backscattered RN16. Our experiments reveal that strictly sticking to $\tau$ would induce too many randomly distributed symbols in the I-Q plane, which make it very difficult for the reader to do proper symbol clustering.

We tackle the above issue by introducing a short \emph{phase-discovery} period lasting $\gamma$ that must satisfy two requirements. First, it starts from a random hopping interval hard to predict by the adversary. Second, it covers at least one phase inversion in the FM0 symbols of the RN16 message. An RN16 message comprises a 6-bit preamble, a 16-bit random number, and one dummy bit. According to FM0 encoding in Fig.~\ref{fig:fm0_baseband}, there is a phase inversion at every symbol boundary and also one in the middle of each data-0 symbol, but the FM0 preamble contains a phase-inversion violation at the fifth symbol labeled ``v''. So the longest time that the RFID card does not invert the signal phase is  $1.5 T_{\textrm{pri}}$. Since the reader does not know when backscattering (i.e., the RN16 transmission) starts, we set $\gamma=2T_{\textrm{pri}}=10\tau$ to satisfy both requirements above. The phase-discovery period obviously consists of 10 hopping intervals. In addition, the reader uses the same CW phase in the odd-numbered hopping intervals and performs random phase hopping in the rest intervals of the phase-discovery period.

Now we explain the protocol details with the timing diagram in Fig.~\ref{fig:jam_protocol}. After sending the Query message, the RFID reader starts the phase-hopping duration at $T_1'$ which is divided into short hopping intervals of $\tau=5\mics$ long. We require the phase-hopping duration to at least cover the range $[244\mics, 247\mics + T_{\textrm{RN16}}]$, where $T_{\textrm{RN16}}=575\mics$ \cite{UHFG218}. So we set $T_1'=240\mics$ and  the phase-hopping duration to $600\mics$ long which corresponds to 120 hopping intervals. For each rhythm-query round, the reader determines 24 CW phase values 
\begin{equation}\label{eq:phase_lib}
\Theta = [\theta_{\textrm{init}}, \theta_{\textrm{init}} + 1, \theta_{\textrm{init}} +2, \dots, \theta_{\textrm{init}} + 23],
\end{equation}  
where $\theta_{\textrm{init}}$ is a random integer from $[0, 360)$. Assume that the phase-recovery period starts at $T_{1}' + n\tau$, where $n\in[0,110]$ is randomly chosen by the reader because the phase-hopping duration lasts 120 hopping intervals. In addition, the reader randomly selects $\theta_{\textrm{reserve}}\in \Theta$ and uses it for the five odd-numbered hopping intervals (represented by lined blocks) in the phase-recovery period. Finally, the reader performs random phase hopping across the remaining 23 phase values in the rest 115 hopping intervals (represented by gray blocks) such that each phase value in $\Theta$ (including $\theta_{\textrm{reserve}}$) is used exactly five times in each rhythm-query round. 

Fig.~\ref{fig:hoppingEffect} gives an example for the efficacy of our protocol, which is based on our prototyping implementation on a USRP 2954R device.  The phase-hopping duration is from $0.1\s$ to $0.7\s$, and the reader's received signals in the phase-recovery period are enclosed by the black rectangle. Since the reader knows exactly when the phase-recovery period starts, it can precisely locate the symbols associated with the constant phase $\theta_{\textrm{reserve}}$. As shown in Fig.~\ref{fig:jam_sig_tag_const}, the reader can easily cluster these symbols into states S1 and S2 whereby to extract the correct phase of backscattered signals. To highlight the correctness of our protocol, we also show complete phase plots obtained by the reader in Fig.~\ref{fig:dist_phase} with our phase-hopping protocol, which match well with those on a traditional RFID reader without phase hopping \cite{RX420}. In contrast, the adversary does not know when the phase-recover period starts. So it has to exploit all the sniffed symbols for phase recovery, which is almost impossible as shown in Fig.~\ref{fig:jam_sig_tag_const_all}.  

\subsection{Resilient Analysis against Advanced Eavesdropping}\label{sec:adv_eaves}

The proposed phase-hopping protocol can thwart basic eavesdropping attacks in which the adversary has only one sniffer that overhears the superposition of the backscattered signal and CW with random phase hopping. Now we analyze its resilience to advanced eavesdropping attacks in which the adversary has an additional sniffer at distance $d_1$ from the reader and $d_2$ from the card. The adversary can also vary $d_1$ and $d_2$ arbitrarily. Theoretically speaking, the second sniffer also receives the superposition of the backscattered signal and CW with random phase hopping. Assume that the adversary can make $d_2$ large enough such that the backscattered signal is attenuated too much to detect, while keeping $d_1$ sufficiently small such that the CW signal is still strong enough. The signal overheard by the second sniffer thus corresponds to CW alone. 
The adversary can then derive the phase-hopping sequence and correlate it with the signals obtained by the first sniffer to recover the phase information of backscattered signals. 

To analyze the feasibility of the advanced eavesdropping attack above, we assume the free-space path loss (FSPL) model for RFID signal propagation, $	\textrm{FSPL} = (\frac{4\pi d}{\lambda})^2$, 
where $d$ is the distance between antennas, and $\lambda$ is the CW wavelength. Assume that the RFID card is at distance $d_0$ from the reader. The power of the reader's signal at the card is $	P_{\textrm{card}} = P_{t}G_{t}(\frac{\lambda}{4\pi d_{0}})^2$, where $P_{t}$ is the reader's transmission power, $d_{0}$ is the distance between reader, and $G_{t}$ is the reader's antenna gain.
According to \cite{NikitAnt08}, the EIRP (Equivalent Isotropically Radiated Power) of passive RFID cards is
\begin{equation}
\textrm{EIRP}_{\textrm{card}} = P_{\textrm{reader}}\frac{4\pi\sigma}{\lambda^2} = P_{t}G_{t}\frac{\sigma}{4\pi d_{0}^2}\;,
\end{equation}
where $\sigma$ denotes the tag's radar cross section (RCS) \cite{NikitAnt08}. $\sigma$ mainly depends on the impedance of card antenna and chip and depicts the backscattered power strength tag.

The second sniffer receives the superposition of CW and the backscattered signal. The signal strength for CW can be expressed by 
\[
P_{\textrm{CW},d_1}= \frac{P_{t}G_{t}G_{r}}{\rm{FSPL_{reader}}} = P_{t}G_{t}G_{r}(\frac{\lambda}{4\pi d_{1}})^2\;,
\]
where $G_r$ denotes the second sniffer's antenna gain. 
Similarly, the signal strength for the backscattered signal can be expressed by 
\[
P_{\textrm{BS},d_2} = \frac{\rm{EIRP_{card}}G_{r}}{\rm{FSPL_{reader}}} = P_{t}G_{t}G_{r}\frac{\sigma\lambda^2}{(4\pi)^3 d_{0}^2 d_{2}^2}.
\]
Let $\tau_{\textrm{rx}}$ and $\tau_{\textrm{dec}}$ denote the minimum signal strengths that the sniffer can detect and decode RFID signals, respectively. The advanced eavesdropping attack works if and only if $P_{\textrm{CW},d_1}\geq \tau_{\textrm{dec}}$ and $P_{\textrm{BS},d_2}\leq \tau_{\textrm{rx}}$ can simultaneously hold.  It is equivalent for the adversary to find $d_1$ and $d_2$ that satisfy 
\[
d_{1} \leq \sqrt{\frac{P_{t}G_{t}G_{r}}{\tau_{\textrm{dec}}}(\frac{\lambda}{4\pi})^2} 
\]
and
\[
d_{2} \geq \sqrt{\frac{P_{t}G_{t}G_{r}}{\tau_{\textrm{rx}}}\frac{\sigma \lambda^2}{(4\pi)^3d_{0}^{2}}}\;.
\]
The above requirement corresponds a \emph{vulnerable region} outside the circle centered at the card with radius $d_2$ and inside the circle centered at the reader with radius $d_1$. In Section~\ref{sec:evaluation}, we experimentally show that the vulnerable region can be very difficult or infeasible to find in practice. 

\section{Evaluation}\label{sec:evaluation}

\subsection{Experimental Setup}\label{sec:experimentSetup}

We used two Impinj R420 readers (GX21M and USA2M1 models) with Laird S9028 antenna. GX21M does not use frequency hopping, while USA2M1 does. The data from USA2M1 were calibrated with the method in Section~\ref{sec:mitigateFHSS} and then combined with the data from GX21M. We used three types of RFID tags, including SMARTRAC R6 DogBone, Impinj E51, and Alien 9640. In addition, we prototyped the phase-hopping protocol on a USRP 2954R and also used an R\&S FSVR7 real-time spectrum analyzer for signal analysis.

We compared the classification performance of SVM, NN, and CNN. The comparison was based on the SVM toolbox in Matlab and the NN and CNN implementations in PyTorch. We used a fully connected NN with one hidden layer and 256 perceptions. In addition, the CNN we used has two 1D convolutional layers and a kernel size of 2. All the training and classification procedures were performed on a Ubuntu desktop with i7-8700k CPU and 16 GB RAM.  

We recruited 19 volunteers from China and US who are either undergraduate or graduate students. Each volunteer tapped a random RFID tag 40 times according to his/her self-chosen rhythm. 
Most chosen rhythms last $6\second$ to $12\second$ with the average and variance equal to $9.61\second$ and $5.86\s$, respectively. The RFID reader-tag distance was always about 40 inches. We collected 760 tapping rhythm samples in total.

\begin{table}[t]
	\centering
	\caption{Classification accuracy (\%) with legitimate users and brute force attackers.}
	\label{tab:kfoldAccuracy}
	\begin{tabular}{|c|p{0.24in}|p{0.24in}|p{0.24in}|p{0.24in}|p{0.24in}|p{0.24in}|}
		\hline
		& \multicolumn{2}{c|}{SVM}    & \multicolumn{2}{c|}{NN}   & \multicolumn{2}{c|}{CNN} \\ 
		\hline
		$K$ & train  & test & train & test & train & test \\
		\hline 
		4  & 100      & 91.97      & 100      & 91.05     & 98.29       & 85.75      \\
		\hline
		5  & 100      & 92.48      & 100      & 90.44     & 98.13       & 87.92      \\
		\hline
		6  & 100      & 92.27      & 100      & 92.07     & 98.42       & 87.39      \\
		\hline
		7  & 100      & 92.96      & 99.87       & 92.50     & 98.57       & 88.68      \\
		\hline
		8  & 100      & 93.05      & 100      & 92.29     & 98.29       & 91.83      \\
		\hline
		9  & 100      & 92.50      & 100      & 93.45     & 98.16       & 90.26      \\
		\hline
		10 & 100      & 93.82      & 100      & 93.16     & 98.42       & 93.68      \\
		\hline
		15 & 100      & 94.65      & 100      & 94.19     & 98.81       & 94.88      \\
		\hline
		20 & 100      & 95.39      & 100      & 94.21     & 99.47       & 96.58      \\
		\hline
	\end{tabular}
	\vspace{-.15in}
\end{table}

\subsection{Performance Results with Legitimate Users (Resilience to Brute Force Attacks)} 

We first evaluate the performance of RF-Rhythm under the brute force attack. For this evaluation, we randomly choose $K$ rhythm samples from all the 19 volunteers to form a training set of $ 19K $ samples to train a classifier for each volunteer. The remaining rhythm samples are treated as the testing set. We do this evaluation 10 times for each volunteer and report the average result. When the classifier of each volunteer is tested against the data samples of all the other 18 volunteers, it amounts to launching a brute forth attack on RF-Rhythm. 

Table~\ref{tab:kfoldAccuracy} shows the training and testing accuracy with SVM, NN, and CNN classifiers, where (classification) accuracy is defined as the percentage of correct predictions. Overall, RF-Rhythm can admit legitimate users and reject random impostors with overwhelming probability under all three classifiers. In addition, both SVM and NN work very well even when $K$ is very small. A smaller $ K $ means a legitimate user can input his/her tapping rhythm fewer times in the enrollment phase, leading to shorter enrollment time and higher usability. In contrast, CNN needs more training data to outperform SVM and NN in the testing phase, which is anticipated.  

\begin{figure}[t]
	\centering
	\includegraphics[width=2.8in]{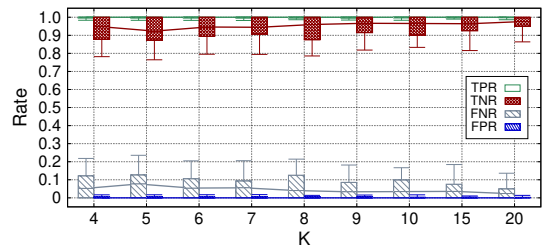}
	\caption{Classification performance of RF-Rhythm using SVM.}
	\label{fig:svm_tfpn}
	\vspace{-.15in}
\end{figure}

Fig.~\ref{fig:svm_tfpn} demonstrates the true-positive rate (TPR), true-negative rate (TNR), false-negative rate (FNR), and false positive rate (FPR) for SVM. Here we use the box plots with the 0th, 25th, 50th, 75th, and 100th percentiles shown. The high classification performance of RF-Rhythm is quite clear. Similar results are obtained for NN and CNN as well and omitted here due to space constraints. 

Since the same user may perform enrollment and authentication at a different distance from the RFID reader, we also evaluate the impact of this distance factor. In this experiment, we place an RFID card at 20, 40, 80, and 120 inches from the RFID reader and let a random volunteer input his tapping rhythm 40 times at each testing location. Then we train a classifier for the volunteer at each location by using his rhythm samples collected there and the samples of all the other 18 volunteers as the training data. Finally, we test each obtained classifier against the volunteer's rhythm samples collected at the same and different locations. Table~\ref{tab:distance_measure} shows the classification accuracy for this evaluation, where E1\&T1, E2\&T2, E3\&T3, and E4\&T4 denote the enrollment and testing locations at 20, 40, 80, and 120 inches, respectively. If the enrollment and testing locations are the same, we randomly divide the volunteer's samples at that location into 2 parts for training and testing, respectively; otherwise, all the 40 samples are used for training in each enrollment location. The results represent the average of 10 runs.  It is clear that RF-Rhythm is robust to enrollment-authentication location variations.

\begin{table}[t]
	\tiny
	\centering
	\caption{Classification accuracy for enrollment-authentication location variations.}
	\label{tab:distance_measure}
	\begin{tabular}{|c|p{0.2cm}|p{0.2cm}|p{0.2cm}|p{0.2cm}|p{0.2cm}|p{0.2cm}|p{0.2cm}|p{0.2cm}|p{0.2cm}|p{0.2cm}|p{0.2cm}|p{0.2cm}|}
		\hline
		& \multicolumn{4}{c|}{SVM}             & \multicolumn{4}{c|}{NN}              & \multicolumn{4}{c|}{CNN}              \\ \hline
		& T1       & T2       & T3       & T4      & T1       & T2       & T3       & T4      & T1       & T2       & T3       & T4       \\ \hline
		E1 & 1.0 & 0.925  & 0.8  & 0.9 & 1.0 & 0.925  & 0.8  & 0.9 & 1.0 & 0.975  & 0.9  & 0.95  \\ \hline
		E2 & 1.0 & 1.0 & 1.0 & 0.925 & 1.0 & 1.0 & 1.0 & 0.95 & 1.0 & 1.0 & 1.0 & 0.95  \\ \hline
		E3 & 0.95  & 1.0 & 0.92  & 0.925 & 1.0 & 0.95  & 1.0 & 0.95 & 1.0 & 1.0 & 1.0 & 0.95  \\ \hline
		E4 & 0.95  & 1.0 & 1.0 & 0.92 & 1.0 & 1.0 & 0.95  & 0.95 & 1.0 & 1.0 & 1.0 & 1.0 \\ \hline
	\end{tabular}
	\vspace{-.15in}
\end{table}

\subsection{Resilience to Visual Eavesdropping}

We also evaluate the resilience of RF-Rhythm to visual eavesdropping. In this evaluation, we use a high-definition smartphone to video-record each volunteer's entire rhythm-tapping process. Then we recruit five volunteers that act as attackers to watch all the 19 videos and then emulate the tapping rhythms they observe. We consider two scenarios. First, each attacker has a one-time watching of each video and then tries to perform the observed rhythm once. This scenario emulates the shoulder-surfing attack. Second, each attacker can watch each video as many times as they want and then performs each perceived rhythm four times. This scenario emulates the video-taping attack via a spy camera. We totally collect 475 attack samples. Then we build a classifier for each of the 19 volunteers with all the aforementioned 760 rhythm samples as the training data. Finally, we test each attack sample with the corresponding volunteer's classifer.

Table~\ref{tab:attack_performance} shows the rejection rate for visual eavesdroppers, which represents the average of 10 runs. We can see that RF-Rhythm has strong resilience to visual eavesdroppers under all three classification methods. In addition, a visual eavesdropper can intuitively achieve a higher success rate with more observations and authentication attempts. RF-Rhythm can rate-limit unsuccessful authentication attempts to provide a stronger defense. 
\begin{table}[t]
	\caption{Rejection rate (\%) for visual eavesdroppers.}
	\label{tab:attack_performance}
	\centering
	\begin{tabular}{|c|c|c|c|}
		\hline
		& SVM & NN      & CNN     \\ \hline
		one observation, one try & 94.74        & 94.63  & 96.32  \\ \hline
		arbitrary observations, 4 tries & 93.42         & 93.53  & 93.87  \\ \hline
	\end{tabular}
	\vspace{-.15in}
\end{table}

\subsection{Resilience to Basic Rhythm Eavesdropping}

Next we examine the efficacy of our phase-hopping protocol to a rhythm eavesdropper with a single sniffer.  
As shown in Fig.~\ref{fig:jam_sig_tag_const_all}, the adversary can roughly cluster sniffed symbols into states S1 and S2, respectively. But it cannot precisely find the matching S1 and S2 symbols of the same CW phase. We assume that the adversary is very powerful and knows how our phase-hopping protocol works. Since the CW phase in each query round takes random values in $\Theta = [\theta_{\textrm{init}}, \theta_{\textrm{init}} + 1, \theta_{\textrm{init}} +2, \dots, \theta_{\textrm{init}} + 23]$, we wwwassume that the adversary can estimate a candidate phase vector $\Theta'$ from sniffed S1 symbols. Due to noise, interference, and processing errors, $\Theta'$ may overlap but is usually much larger than $\Theta$. The symbols in $\Theta'$ can be much fewer than sniffed S1 symbols. Then the adversary picks an arbitrary sniffed S2 symbol, denoted by $s_2$, and uses each S1 symbol in $\Theta'$ as a candidate matching symbol for $s_2$ to derive a candidate phase of the backscattered RN16. The probability of a correct guess  is simply $1/|\Theta'|$. Each rhythm-query round is about $2.179\ms$ long, and the average tapping-rhythm duration is $9.61\s$ in our experiments. So we need about 4,410 rounds to cover and detect an average tapping rhythm. The probability that the adversary can recover the correct tapping rhythm from sniffed signals can be estimated by $\tilde{P}=(1/|\Theta'|)^n$. For example, if $|\Theta'|= 24|48|72$, the adversary can succeed with negligible probability. Therefore, our phase-hopping protocol is highly effective against the basic rhythm-eavesdropping attack. 

\subsection{Resilience to Advanced Rhythm Eavesdropping}

We also evaluate the resilience of RF-Rhythm to advanced rhythm-eavesdropping attacks in which the adversary has two sniffers at strategic locations. In Section~\ref{sec:adv_eaves}, we identify a theoretical vulnerable region in which this attack can succeed. In this section, we show that the vulnerable region may not be easily found by an adversary with reasonable equipment. 

In this evaluation, we assume that the adversary places his second sniffer $d_1$ from the RFID reader and $d_2$ from the RFID card. For simplicity, we assume that the reader, tag, and sniffer are on the straight line. This is a reasonable assumption because most commonly used RFID antennas are directional with a relatively focused and narrow radio wave beam. We implement a EPC Gen2 RFID reader prototype \cite{KargaFul15} on an NI USRP 2954R and assume that the adversary has a similar sniffer device. We also use an R\&S FSVR7 real-time spectrum analyzer for signal measurements. Recall that $\tau_{\textrm{rx}}$ and $\tau_{\textrm{dec}}$ denote the minimum signal strengths that the sniffer can detect and decode RFID signals, respectively. According to our measurements, $\tau_{\textrm{rx}} = -81.21 \deci\bel\m$ and $\tau_{\textrm{dec}} = -55.98 \deci\bel\m$. 

To emulate the attack, we vary the RFID card-reader distance $d_0$ from 10 to 40, 80, and 120 inches. For each $d_0$ value, we measure the CW signal strength $P_{\textrm{CW},d_1}$ and the backscattered signal strength $P_{\textrm{BS},d_2}$ at $d_2=40$ inches from the RFID card, which also corresponds to $d_1=d_0+40$ inches. This location is regarded as the sniffer's initial location. The results are shown in Table~\ref{tab:power_res}. Since we assume the reader-card-sniffer line topology, $P_{\textrm{CW},d_1}$ and $P_{\textrm{BS},d_2}$ are attenuated by the same amount when $d_2$ and equivalently $d_1$ increase.  According to our analysis in Section~\ref{sec:adv_eaves}, the advanced eavesdropping attack succeeds if and only if $P_{\textrm{CW},d_1}\geq \tau_{\textrm{dec}}$ and $P_{\textrm{BS},d_2}\leq \tau_{\textrm{rx}}$ can simultaneously hold.  This requires $P_{\textrm{CW},d_1}-P_{\textrm{BS},d_2}\geq \tau_{\textrm{dec}} - \tau_{\textrm{rx}}= 25.23 \decibel\m$ per our measurements. This requirement cannot be satisfied according to Table~\ref{tab:power_res}, so the advanced eavesdropping attack would fail. 

It is possible that a more capable adversary with advanced equipment can successfully overhear the legitimate user's tapping rhythm. Instead of being a perfect solution, RF-Rhythm, however, just aims to enhance the security of a traditional RFID authentication system that is naturally vulnerable to lost/stolen/cloned RFID cards. In other words, RF-Rhythm significantly raises the bar for launching successful attacks on RFID authentication systems.

\begin{table}[t]
	\caption{Power measurements for advanced rhythm eavesdropping.}
	\label{tab:power_res}
	\centering
	\begin{tabular}{|c|c|c|c|}
		\hline
		$d_{0}$/inch & $P_{\textrm{CW},d_1}/\deci\bel\m$ & $P_{\textrm{BS},d_2}/\deci\bel\m$ & $P_{\textrm{CW},d_1} - P_{\textrm{BS},d_2}/\deci\bel\m$ \\ \hline
		10            & -3.30 & -27.91 & 24.61 \\\hline
		40            & -7.98 & -27.00 & 20.78 \\\hline
		80            & -10.15 & -24.02 & 14.53 \\\hline
		120           & -14.52 & -26.43 & 12.29 \\\hline
	\end{tabular}
	\vspace{-.15in}
\end{table}

\subsection{Additional Results}

We also evaluate the computational latency of RF-Rhythm. Our results show that the classifier training can be done in a few seconds, and each tapping rhythm can be classified in less than $1\ms$. In addition, we use a questionnaire to confirm the high usability of RF-Rhythm. These results are omitted here due to space constraints.

\section{Related Work}\label{sec:related}

Rhythm-based authentication for mobile devices has been explored. RhyAuth \cite{ChenYou15} is a two-factor rhythm-based authentication scheme for multi-touch mobile devices. It requires a user to perform a sequence of rhythmic taps/slides on a device screen to unlock the device. In the follow-on work, Beat-PIN \cite{HutchBea18} requires a user to tap the screen of a smartwatch to unlock it. RF-Rhythm differs significantly from RhyAuth and Beat-PIN in the application context, totally different rhythm-extract techniques, adversary models, and countermeasures.  

There is also significant effort on RFID security. For example, novel cryptographic RFID authentication protocols are presented in \cite{kulseLig10, LiPri12,YangAna17}. Haitham \cite{HassaSec15} proposes RF-Cloak to prevent eavesdropping attacks by randomizing the modulation and channel. Selective jamming is proposed in \cite{DingPre18} to prevent unauthorized inquiries to RFID tags. Zanetti and Danev \cite{ZanetPhy10} explore the time interval error, average baseband power and spectral features to fingerprint RFID tags. TapPrint \cite{YangAnt15} uses the phase of backscattered signals combined with the geometric relationship to fingerprint RFID tags. Hu-Fu \cite{WangTow18} uses the inductive coupling of two tags to fingerprint them.
RF-Mehndi \cite{ZhaoRfm19} identifies an RFID card and its user simultaneously by exploring the backscattered signal changes induced by the user's fingertip on a specially build passive tag array. RF-Rhythm explores COTS RFID tags and is complimentary to the above work. 

The phase information of backscattered RFID signals has been explored in many applications, such as gesture recognition \cite{WangMul18, BuRfd18}, action recognition \cite{WangRfk18, JinTow17}, orientation tracking \cite{WeiGyr16}, mechanical features sensing \cite{YangMak16, JinWis18}, and localization \cite{MaMin17}. RF-Rhythm is the first work to extract a tapping rhythm from backscattered RFID signals and is orthogonal to the above work.

\section*{Acknowledgment}
This work was supported in part by the US National Science Foundation under grants CNS-1514381, CNS-1619251, CNS-1651954 (CAREER), CNS-1700039, CNS-1718078, CNS-1824355, CNS-1933047, and CNS-1933069.

%\clearpage
\bibliography{cite,wins,winsPub} 

\begin{thebibliography}{10}
\providecommand{\url}[1]{#1}
\csname url@samestyle\endcsname
\providecommand{\newblock}{\relax}
\providecommand{\bibinfo}[2]{#2}
\providecommand{\BIBentrySTDinterwordspacing}{\spaceskip=0pt\relax}
\providecommand{\BIBentryALTinterwordstretchfactor}{4}
\providecommand{\BIBentryALTinterwordspacing}{\spaceskip=\fontdimen2\font plus
\BIBentryALTinterwordstretchfactor\fontdimen3\font minus
  \fontdimen4\font\relax}
\providecommand{\BIBforeignlanguage}[2]{{%
\expandafter\ifx\csname l@#1\endcsname\relax
\typeout{** WARNING: IEEEtran.bst: No hyphenation pattern has been}%
\typeout{** loaded for the language `#1'. Using the pattern for}%
\typeout{** the default language instead.}%
\else
\language=\csname l@#1\endcsname
\fi
#2}}
\providecommand{\BIBdecl}{\relax}
\BIBdecl

\bibitem{RFID2FA}
\BIBentryALTinterwordspacing
``{Two-Factor Authentication (2FA) Explained: RFID Access Control}.'' [Online].
  Available:
  \url{https://blog.identityautomation.com/two-factor-authentication-2fa-explained-rfid-access-control}
\BIBentrySTDinterwordspacing

\bibitem{Duo}
\BIBentryALTinterwordspacing
``Duo push.'' [Online]. Available:
  \url{https://www.duosecurity.com/product/methods/duo-mobile}
\BIBentrySTDinterwordspacing

\bibitem{ChenYou15}
Y.~Chen, J.~Sun, R.~Zhang, and Y.~Zhang, ``Your song your way: Rhythm-based
  two-factor authentication for multi-touch mobile devices,'' in \emph{IEEE
  INFOCOM}, Hong Kong, China, April 2015.

\bibitem{HutchBea18}
B.~Hutchins, A.~Reddy, W.~Jin, M.~Zhou, M.~Li, and L.~Yang, ``Beat-pin: A user
  authentication mechanism for wearable devices through secret beats,'' in
  \emph{ACM ASIACCS}, Incheon, Republic of Korea, June 2018.

\bibitem{UHFG218}
\BIBentryALTinterwordspacing
``{EPC UHF Gen2 Air Interface Protocol}.'' [Online]. Available:
  \url{https://www.gs1.org/standards/epc-rfid/uhf-air-interface-protocol}
\BIBentrySTDinterwordspacing

\bibitem{RX420}
\BIBentryALTinterwordspacing
``{SPEEDWAY R420 RAIN RFID READER}.'' [Online]. Available:
  \url{https://www.impinj.com/platform/connectivity/speedway-r420}
\BIBentrySTDinterwordspacing

\bibitem{LLUD13}
\BIBentryALTinterwordspacing
``{Speedway Revolution Reader Application Note: Low Level User Data Support}.''
  [Online]. Available:
  \url{https://support.impinj.com/hc/en-us/articles/202755318-Application-Note-Low-Level-User-Data-Support}
\BIBentrySTDinterwordspacing

\bibitem{VemulHum14}
R.~Vemulapalli, F.~Arrate, and R.~Chellappa, ``Human action recognition by
  representing 3d skeletons as points in a lie group,'' in \emph{IEEE CVPR},
  Columbus, OH, June 2014.

\bibitem{WangMov16}
C.~Wang, L.~Xie, W.~Wang, T.~Xue, and S.~Lu, ``Moving tag detection via
  physical layer analysis for large-scale rfid systems,'' in \emph{IEEE
  INFOCOM}, San Francisco, CA, April 2016.

\bibitem{NikitAnt08}
P.~Nikitin and S.~Rao, ``Antennas and propagation in uhf rfid systems,'' in
  \emph{IEEE RFID}, Las Vegas, NV, April 2008.

\bibitem{KargaFul15}
N.~Kargas, F.~Mavromatis, and A.~Bletsas, ``Fully-coherent reader with
  commodity sdr for gen2 fm0 and computational rfid,'' \emph{IEEE Wireless
  Communications Letters}, pp. 617--620, 2015.

\bibitem{kulseLig10}
L.~Kulseng, Z.~Yu, Y.~Wei, and Y.~Guan, ``Lightweight mutual authentication and
  ownership transfer for rfid systems,'' in \emph{IEEE INFOCOM}, Pisa, Italy,
  March 2010.

\bibitem{LiPri12}
T.~Li, W.~Luo, Z.~Mo, and S.~Chen, ``Privacy-preserving rfid authentication
  based on cryptographical encoding,'' in \emph{INFOCOM}, Orlando, USA, March
  2012.

\bibitem{YangAna17}
L.~Yang, Q.~Lin, C.~Duan, and Z.~An, ``Analog on-tag hashing: Towards selective
  reading as hash primitives in gen2 rfid systems,'' in \emph{ACM Mobicom},
  Snowbird, Utah, October 2017.

\bibitem{HassaSec15}
H.~Hassanieh, J.~Wang, D.~Katabi, and T.~Kohno, ``Securing rfids by randomizing
  the modulation and channel,'' in \emph{NSDI}, Oakland, CA, May 2015.

\bibitem{DingPre18}
H.~Ding, J.~Han, Y.~Zhang, F.~Xiao, W.~Xi, G.~Wang, and Z.~Jiang, ``Preventing
  unauthorized access on passive tags,'' in \emph{IEEE INFOCOM}, Honolulu, HI,
  April 2018.

\bibitem{ZanetPhy10}
D.~Zanetti and B.~Danev, ``Physical-layer identification of uhf rfid tags,'' in
  \emph{ACM Mobicom}, Chicago, Illinois, September 2010.

\bibitem{YangAnt15}
L.~Yang, P.~Peng, F.~Dang, C.~Wang, X.~Li, and Y.~Liu, ``Anti-counterfeiting
  via federated rfid tags' fingerprints and geometric relationships,'' in
  \emph{IEEE INFOCOM}, Kowloon, Hong Kong, April 2015.

\bibitem{WangTow18}
G.~Wang, H.~Cai, C.~Qian, J.~Han, X.~Li, H.~Ding, and J.~Zhao, ``Towards
  replay-resilient rfid authentication,'' in \emph{ACM Mobicom}, New Delhi,
  India, October 2018.

\bibitem{ZhaoRfm19}
C.~Zhao, Z.~Li, T.~Liu, H.~Ding, J.~Han, W.~Xi, and R.~Gui, ``Rf-mehndi: A
  fingertip profiled rf identifier,'' in \emph{IEEE INFOCOM}, Paris, France,
  April 2019.

\bibitem{WangMul18}
C.~Wang, J.~Liu, Y.~Chen, H.~Liu, L.~Xie, W.~Wang, B.~He, and S.~Lu,
  ``Multi-touch in the air: Device-free finger tracking and gesture recognition
  via cots rfid,'' in \emph{IEEE INFOCOM}, Honolulu, HI, April 2018.

\bibitem{BuRfd18}
Y.~Bu, L.~Xie, Y.~Gong, C.~Wang, L.~Yang, J.~Liu, and S.~Lu, ``Rf-dial: an
  rfid-based 2d human-computer interaction via tag array,'' in \emph{IEEE
  INFOCOM}, Honolulu, HI, April 2018.

\bibitem{WangRfk18}
C.~Wang, J.~Liu, Y.~Chen, L.~Xie, H.~Liu, and S.~Lu, ``Rf-kinect: A wearable
  rfid-based approach towards 3d body movement tracking,'' in \emph{ACM
  UBICOMP}, Singapore, October 2018.

\bibitem{JinTow17}
H.~Jin, Z.~Yang, S.~Kumar, and J.~Hong, ``Towards wearable everyday body-frame
  tracking using passive rfids,'' in \emph{ACM UBICOMP}, Singapore, October
  2018.

\bibitem{WeiGyr16}
T.~Wei and X.~Zhang, ``Gyro in the air: tracking 3d orientation of batteryless
  internet-of-things,'' in \emph{ACM Mobicom}, New York City, New York, October
  2016.

\bibitem{YangMak16}
L.~Yang, Y.~Li, Q.~Lin, X.-Y. Li, and Y.~Liu, ``Making sense of mechanical
  vibration period with sub-millisecond accuracy using backscatter signals,''
  in \emph{ACM Mobicom}, New York City, New York, October 2016.

\bibitem{JinWis18}
H.~Jin, J.~Wang, Z.~Yang, S.~Kumar, and J.~Hong, ``Wish: Towards a wireless
  shape-aware world using passive rfids,'' in \emph{ACM MobiSys}, Munich,
  Germany, June 2018.

\bibitem{MaMin17}
Y.~Ma, N.~Selby, and F.~Adib, ``Minding the billions: Ultra-wideband
  localization for deployed rfid tags,'' in \emph{ACM Mobicom}, Snowbird, Utah,
  October 2017.

\end{thebibliography}
\bibliographystyle{IEEEtran}

\end{document}